\documentclass[aps,preprintnumbers,amsmath,amssymb]{revtex4}
\usepackage{subeqnarray}
\usepackage{graphicx}
\usepackage{color}
\definecolor{navyblue}{rgb}{0.3,0.3,1}
\definecolor{purple}{rgb}{0.6,0,0.5}
\usepackage[colorlinks=true, pdfstartview=FitV, linkcolor=purple, citecolor=
blue,urlcolor=navyblue]{hyperref}

\begin{document}

\title{$B^+\to K^-\pi^+\pi^+$: three-body final state interactions and $K\pi$ isospin states}
 \date{\today}

\author{J. H. Alvarenga Nogueira$^a$, T. Frederico$^a$, O. Louren\c{c}o$^b$}

\affiliation{$^a$Instituto Tecnol\'ogico de Aeron\'autica, 12228-900, S\~ao Jos\'e dos Campos, SP, Brazil  \\
$^b$Universidade Federal do Rio de Janeiro, 27930-560, Maca\'e, RJ, Brazil}

\begin{abstract}
Final state interactions are considered to formulate the $B$ meson decay amplitude for the 
$K\pi\pi$ channel. The Faddeev decomposition of the Bethe-Salpeter equation is used  in 
order to build a relativistic three-body model within the light-front framework. The 
S-wave scattering amplitude for the $K\pi$ system is considered in the $1/2$ and $3/2$ 
isospin channels with the set of inhomogeneous integral equations solved perturbatively. 
In comparison with previous results for the $D$ meson decay in the same channel, one has 
to consider the different partonic processes, which build the source amplitudes, and the 
larger absorption to other decay channels appears, that are important features to be 
addressed. As in the $D$ decay case, the convergence of the rescattering perturbative 
series is also achieved with two-loop contributions.
\end{abstract}


\maketitle
\section{Introduction}

Heavy quark decays are largely explored in the literature. Due to the large $B$ 
meson mass ($m_b$), there are several approaches for $B$ decays based on QCD effective 
field theories within heavy quark expansions~\cite{QCDF,pQCD,SCET}. They are based on 
factorization of the hadronic matrix elements and mainly consider short-distance physics. 
The weak effective Hamiltonian is constructed based  on tools from quantum field 
theory, such as the operator product expansion to separate the problem in the 
long-distance and short-distance physics. The perturbative treatment is justified by the 
fact that the strong coupling constant $\alpha_s$ is small in high energy short-distance 
processes. The long-distance physics and its non-perturbative nature leads to divergent 
amplitudes that are complicated to deal with and requires care. The called soft Final 
State Interactions (FSI) shows to be essential in studies involving $B$ meson decays, 
since it does not disappear for large $m_b$~\cite{DonoghuePRL}. Contributions coming from 
long-distance inelastic rescattering is expected to be the main source of soft FSI and can 
be substantial in Charge-Parity ($CP$) violation distributions~\cite{ 
CPT,Nogueira15}. Rescattering effects can also explain the appearance of events in 
very suppressed decay channels. A recent experimental study of the charmless $B_c$ decay 
to the $KK\pi$ channel, which within the Standard Model can only occur by weak 
annihilation diagrams, shows some events in the phase space of this 
channel~\cite{LHCbBc}. This can be related with hadronic rescattering inelastic 
transitions to that final decay channel. QCD factorization calculations of two-body $B_c$ 
decays, also suppressed, can explain that small branching ratios~\cite{QCDFBc}. 
Contributions coming from final state interaction for the $B^+ \to J/ \psi \pi^+$ decay 
within the QCD factorization approach was further considered in~\cite{QVDFJpsi}.

FSI play an important role in heavy meson weak decays. As a test of 
$CP$ violation, FSI are essential to guarantee $CPT$ invariance~\cite{CPT,Nogueira15}. A 
practical theoretical approach was used to study these three-body charmless $B^\pm$ decays 
in~\cite{CPT}. A more general formulation, including resonances and its interferences, 
applied for four $B$ decay channels is found in Ref.~\cite{Nogueira15}. $CP$ violation in 
the low invariant mass of the $\pi\pi$ system of the $B \to \pi \pi \pi$ decay channel is 
also studied in Ref.~\cite{zhang1}, where contributions from scalar and vector 
resonances are considered. The S-wave $\pi\pi$ elastic scattering in the region below the 
$\rho$ mass has also the important contribution from the $f_0(600)$ resonance, as showed 
in Ref.~\cite{kang} for a four-body semileptonic decay.

Our goal in the present work is to address the issue of three-body FSI in the  
specific $B^{+}\rightarrow K^{-}\pi^{+}\pi^{+}$ decay, with emphasis in the S-wave 
$K^{-}\pi^{+}$ amplitude. In order to proceed in such direction, we closely follow 
the formalism developed for the $D$ decay in Ref.~\cite{JHEP14}. Our study is based in a 
relativistic model for the three-body FSI that was applied to the $D^{+}\rightarrow 
K^{-}\pi^{+}\pi^{+}$ decay~\cite{JHEP14,karinnpb,MagPRD11}. In Ref.~\cite{JHEP14}, the 
isospin projection of the decay amplitude was performed to study different isospin state 
contributions to the $K^{-}\pi^{+}$ rescattering. In that model, by starting from a  
Bethe-Salpeter like equation and using the Faddeev decomposition, the decay amplitude  was 
separated into a smooth term and a three-body fully interacting contribution. Moreover, 
the amplitude was factorized in the standard two-meson resonant amplitude times a reduced 
complex amplitude for the bachelor meson, that carries the effect of the three-body 
rescattering mechanism. The off-shell bachelor amplitude is a solution of an inhomogeneous 
Faddeev type integral equation, that has as input the S-wave isospin $1/2$ and $3/2$ 
$K^{-}\pi^{+}$ transition matrix. In the Faddeev formulation, the integral equation has a 
connected kernel, which is written in terms of the two-body amplitude. The light-front 
(LF) projection of the equations~\cite{sales00} was performed to simplify the numerical 
calculations, and interactions between identical charged pions were neglected. A different 
coupled-channel framework, considering both $\pi\pi$ and $K\pi$ empirical scattering 
amplitudes, was used in Ref.~\cite{nakamura} to study the $D^+ \to K^- \pi^+ \pi^+$ Dalitz 
plot.

Here we discuss the perturbative solutions of the LF integral equations for the bachelor 
amplitude in the $B$ meson decay. To check the convergence of the series expansion, we go 
up to terms of third order in the two-body transition matrix. The numerical results 
for the $B^{+}\rightarrow K^{-}\pi^{+}\pi^{+}$ decay with three-body FSI and $K\pi$
interactions in $I=1/2$ and $3/2$ states are presented. The S-wave $K\pi$ amplitude 
depends on the isospin and orbital angular momentum of the system. There are two isospin 
states possible for this system, namely, $I=1/2$ and $I=3/2$. The LASS experimental
data~\cite{LASS} shows resonances and the corresponding scattering amplitude poles only in 
the isospin $1/2$ channel. This feature is used here to model the $K\pi$ S-matrix used in 
the $B$ decay amplitude.

\section{Decay amplitude for $B^+\to K^-\pi^+\pi^+$ decay with FSI}
\label{sec:FSIcov}

\subsection{S-wave $K\pi$ scattering amplitude}
\label{sec:kpiampl}

The three-body rescattering model used here to study the decay amplitude with FSI, 
requires a two-body transition matrix as input. In the same way we have done in 
the $D^+\to K^-\pi^+\pi^+$ decay in Ref.~\cite{JHEP14}, the $K\pi$ S-wave elastic 
scattering amplitude is introduced in the resonant $I_{K\pi}=1/2$ and non-resonant 
$I_{K\pi}=3/2$ isospin states. We use the same parametrization fitted to the LASS 
data~\cite{LASS} including two resonances above $K^*_0(1430)$, namely $K(1630)$ and 
$K_0^*(1950)$. The $K(1630)$ is usually considered in $B \to K \pi \pi$ amplitude analysis 
and it is more reliable its presence in that channel. However, the existence of the 
$K_0^*(1950)$ resonance is less clear and is usually included in a more indirect way in 
experimental analysis.
The main reason to use the additional 
$K_0^*(1630)$ and $K_0^*(1950)$ resonances is the LASS data, where the whole kinematical 
range up to $1.89$~GeV is fitted. In our analysis, we neglect the $\pi\pi$ 
interaction. The same approximation was also considered in the $D$ decay case of 
Ref.~\cite{JHEP14}.

The parametrized S-matrix ($S^{\mbox{\tiny{1/2}}}_{K\pi}$) is written as:
\begin{eqnarray}
S^{\mbox{\tiny{1/2}}}_{K\pi}= \frac{k\cot\delta+i \,k}{k\cot\delta-i
\,k}\, \prod_{r=1}^3 \frac{M^2_r-M^2_{K\pi}+i z_r \bar\Gamma_r}
 {M^2_r-M^2_{K\pi}-i z_r\Gamma_r}
\label{skpi12}
\end{eqnarray}
where $z_r=k\,M^2_r/(k_r\,M_{K\pi})$ and $k$ is the c. m. momentum of each meson of the 
$K\pi$ pair. Following this S-matrix, the scattering amplitude reads
\begin{eqnarray}
\tau_{\mbox{\tiny{$I_{K \pi}$}}}\left(M^2_{K\pi}\right)=4\pi\,\frac{M_{K\pi}}{k}
\left(S^{\mbox{\tiny{$I_{K \pi}$}}}_{K\pi}-1\right). \label{tau12}
\end{eqnarray}
The resonance parameters associated to the $K_0^*(1430)$, $K_0^*(1630)$ and $K_0^*(1950)$ resonances are
$(M_r,\Gamma_r,\bar\Gamma_r)$ given by: (1.48,0.25,0.25), (1.67,~0.1,0.1) and (1.9,~0.2,~0.14), respectively.

The non-resonant part of the scattering amplitude is parameterized by an effective range expansion as
$k\cot\delta=\frac{1}{a}+\frac12 r_0\, k^2$ using $a=1.6$~GeV$^{-1}$ and $r_0=3.32$~GeV$^{-1}$.
By using such a model, the S-wave $K\pi$ amplitude in the $I=3/2$ state is given by
$S^{\mbox{\tiny{3/2}}}_{K\pi}= \frac{k\cot\delta+i \,k}{k\cot\delta-i
\,k}$,
with the effective range expansion parameters $a=-1.00$~GeV$^{-1}$ and $r_0=-1.76$~GeV$^{-1}$
taken from Ref.~\cite{estabrooks}.

The parametrization from the three-resonance model and the $I_{K\pi}=1/2$ S-wave 
phase-shift compared to the LASS data shows good agreement. The results of the 
parametrization for  $|S^{\mbox{\tiny{1/2}}}_{K\pi}-1|/2$ are shown and discussed in more 
details in Ref.~\cite{JHEP14}. 

\subsection{Three-body rescattering Bethe-Salpeter model}

The full decay amplitude including the rescattering series and the $3\to
3$ transition matrix is written as \cite{JHEP14}:

\begin{eqnarray}
{\cal A}(k_{\pi},k_{\pi^\prime})=B_0(k_{\pi},k_{\pi^\prime}
)
+\int \frac{d^4q_\pi d^4q_{\pi^\prime}}{(2\pi )^8}T(k_{\pi},
k_{\pi^\prime};q_{\pi},q_{\pi^\prime})S_\pi(q_\pi)\,S_{\pi}(q_{\pi^\prime})S_K(K-q_{\pi^\prime}-q_{\pi})
B_0(q_\pi,q_{\pi^\prime}) \ ,\label{D}
\end{eqnarray}
where the momentum of the pions  are
$k_\pi$ and $k_{\pi^\prime}$.

The short-distance physics resides in the $B_0(k_\pi,k_{\pi^\prime})$ amplitude, which 
represents the quark level amplitude. The sum of rescattering diagrams, considered in the 
ladder approximation, is in the second term of Eq. (\ref{D}) and composes the long range 
physics. This term is composed by the $3\to 3$ transition matrix 
$T(k_{\pi},k_{\pi^\prime};q_{\pi},q_{\pi^\prime})$ with the source term and the meson 
propagators $S_i(q_i)=i(q_i^2-m_i^2+i \epsilon)^{-1}$, where self-energies are neglected. 
The $K\pi$ transition matrix sum all $2\to 2$ collision terms. The full transition matrix 
with the FSI is a solution of the Bethe-Salpeter equation, used with its Faddeev 
decomposition.

\subsection{Decay amplitude}

The full three-body T-matrix gives the final state interactions between the mesons in the 
decay channel and it is a solution of the Bethe-Salpeter equation. Here we follow the 
formalism developed in Ref.~\cite{JHEP14}, where the Faddeev decomposition including only 
two-body irreducible diagrams for spinless particles without self-energies is considered. 
Only two body interactions are considered, involving all three-particles except between 
the equal charged pions.

The two-body transition matrix written with a four-conservation delta factorized out reads
\begin{eqnarray}
T_{i}(k'_j,k'_k;k_j,k_k)=(2\pi)^4\tau_i(s_{i}) \,S^{-1}_i(k_i)\,\delta(k'_i-k_i)
\ , \label{eq42}
\end{eqnarray}
where the Mandelstam variable $s_{i}=(k_j+k_k)^2$ is the only dependence considered 
and $\tau_i(s_{i})$ is the unitary S-wave scattering amplitude of particles $i$ and $j$. 
Using the separable form of Eq.~\ref{eq42} the problem is reduced to a four-dimensional 
integral equation in one momentum variable for the Faddeev components of the vertex 
function.

The full decay amplitude considering interactions between all the final states mesons reduces to 
\begin{eqnarray}
{\mathcal A}_0(k_i,k_j)=B_0(k_i,k_j)+\sum_\alpha\tau(s_{\alpha})\xi^\alpha(k_\alpha) \ ,
\label{eq48}
\end{eqnarray}
where the subindex in ${\cal A}_0$ denotes the s-wave two-meson scattering and the 
bachelor amplitude $\xi(k_i)$ carries the three-body rescattering effect and is 
represented by the connected Faddeev-like equations
\begin{multline}
\xi^i(k_i)=\xi^i_0(k_i)+\int \frac{d^4q_j }{(2\pi )^4}S_j(q_j)S_k(K-k_i-q_k)\tau_j(s_{j})\xi^j(q_j) 
+\int \frac{d^4q_k }{(2\pi )^4} S_j(K-k_i-q_k)S_k(q_k)\tau_k(s_{k})\xi^k(q_k) \ .
\label{eq47}
\end{multline}
with $q_k=K-k_i-q_j$. Eq.~\ref{eq47} both amplitude and phase depending on the bachelor 
meson on-mass-shell momentum and $\tau(s_{i})$ can takes into account two-meson 
resonances. The parameterized $K\pi$ scattering amplitude $\tau_i(M^2_{K\pi})$  
reproduces the LASS experimental~\cite{LASS} S-wave phase-shift in the isospin $1/2$ and 
$3/2$ channels.

By taking into account all the model assumptions, the decay amplitude for the 
$B^+\to K^-\pi^+\pi^+$ process is given by
\begin{align}
{\mathcal A}_0(k_{\pi},k_{\pi^\prime})=B_0(k_{\pi},k_{\pi^\prime})+
\tau(M^2_{K\pi})\xi(k_{\pi^\prime})
+\tau(M^2_{K\pi^\prime})\xi(k_{\pi}) \ ,
\label{deq2}
\end{align}
where $M^2_{K\pi}= (K-k_{\pi^\prime})^2$, $M^2_{K\pi^\prime}= (K- k_{\pi})^2$ and the 
bachelor pion on-mass-shell momentum is given by
\begin{eqnarray}
|{\mathbf
k}_\pi|=\left[\left(\frac{M_B^2+m_\pi^2-M^2_{K\pi^\prime}}{2\,M_B}\right)^2
-m_\pi^2\right]^\frac12 \ .
\label{kpi}
\end{eqnarray}
The rescattering series comes from the solution of Eq.~(\ref{deq5}), where the second and 
third terms in Eq. (\ref{eq47}) correspond to higher order loop diagrams.

The inhomogeneous integral equation for the spectator amplitude in the three-body collision process
is a function only of the bachelor momentum (see \cite{JHEP14}),
\begin{align}
\xi(k)=\xi_0(k) +\int\frac{d^4q}{(2\pi)^4}\tau\left((K-q)^2\right)\,S_K(K-k-q)\,S_\pi(q)\,\xi(q),
\label{deq5}
\end{align}
where the first term is
\begin{eqnarray}
\xi_0(k)=\int \frac{d^4q}{(2\pi )^4}S_{\pi}(q)S_K(K-k-q)B_0(k,q),
\label{deq6}
\end{eqnarray}
with the partonic decay amplitude described by $B_0(k,q)$.

The two basic contributions for the decay amplitude are the
well behaved function $B_0(k_{\pi},k_{\pi^\prime})$ and three-body rescattering term 
$\tau\left(M^2_{K\pi^\prime}\right)\xi(k_\pi)$. The operator $\tau$ acts on the isospin 
states $1/2$ and $3/2$. The complex decay amplitude can be decomposed in terms of phase 
and amplitude as
\begin{align}
A(M^2_{K\pi^\prime})=\frac{1}{2} \langle K\pi\pi|B_0\rangle+\langle K\pi\pi|\tau(M^2_{K\pi^\prime})
|\xi(k_{\pi})\rangle = a_0(M^2_{K\pi^\prime})e^{i\Phi_0(M^2_{K\pi^\prime})} ,
\label{kpiampl}
\end{align}
which is a function of only $M^2_{K\pi^\prime}$ and $| K\pi\pi\rangle$ represents
the state in isospin space.

\section{FSI Light-Front equations}
\label{sec:FSILF}

The equations presented for the decay processes considering FSI effects are simplified 
when treated in Light-Front Dynamics. The light-front (LF) projection of the 
four-dimensional coupled equations presents a three-dimensional form. Such 
a technique was successfully applied for the heavy meson decays presented in 
Ref.~\cite{JHEP14} and will also be used here to treat the $B \to K \pi \pi$ 
decay problem. 

After all manipulations, discussed in details in~\cite{JHEP14}, the integral equation in 
terms of the LF variables reads
\begin{eqnarray}
\xi^i(y,\vec k_\perp) = \xi^i_0(y,\vec k_\perp)+ \frac{i}{2(2\pi)^3}\int_0^{1-y} \frac{dx 
}{x(1-x-y)}\int d^2q_{\perp}
\left[\frac{\tau_j\left(M^2_{ik}(x,q_\perp)\right)\xi^j(x,\vec q_\perp)}{M^2-M_{0}^2(x,\vec q_\perp;y,\vec k_\perp)+i\varepsilon}
+ (j\leftrightarrow k)
\right]
, \label{qpt12}
  \end{eqnarray}
where $M^2=K^\mu K_\mu$,  $y=k^+_i/K^+$, $x=q^+_j/K^+$ or $x=q^+_k/K^+$  in the first or 
second integral in the right-hand side of the equation.
The free three-body squared  mass is
\begin{eqnarray}
 M_0^2(x,\vec q_\perp;y,\vec k_\perp)=\frac{k_\perp^2+m^2_i}{y}
 +\frac{q_\perp^2+m^2_j}{x} + \frac{(\vec k_\perp+\vec q_\perp)^2+m^2_k}{1-x-y}.
\label{qpt14}
\end{eqnarray}
The argument of the two-body amplitude $\tau_j\left(M^2_{ik}(x,q_\perp)\right)$ should be
understood as \mbox{
$ M_{ik}^2(x,q_\bot) = (1-x)\left(M^2 - \frac{q_\bot^2 +
m_j^2}{x}\right) - q_\bot^2 \ .$
}
The driven term in Eq.~(\ref{qpt12}) is rewritten as
\begin{eqnarray}
\xi^i_0(y,\vec k_\perp)= \frac{i}{2(2\pi)^3} \int_0^{1-y} \frac{dx}{x(1-y-x)}\int d^2q_{\perp}
\frac{B_0(x,\vec q_\perp;y,\vec k_\perp)}{M^2-M_0^2(x,\vec q_\perp;y,\vec k_\perp)+i\varepsilon}
 . \label{qpt13}
\end{eqnarray}
Since the integral over $q_\perp$ is divergent, a regularization procedure is 
needed. Here we use a finite subtraction constant $\lambda(\mu^2)$, and a 
subtraction point within the integration kernel of Eq.(\ref{qpt13}). This method 
leads to the following driven term
\begin{eqnarray}
\xi_0(y,k_\bot)=\lambda(\mu^2)+\frac{i}{2 }\int_0^1\frac{dx}{x(1-x)}\int_0^{2\pi}
d\theta\int_0^\infty \frac{dq_\bot q_\bot}{(2\pi)^3} \left[\frac{1}{M_{K\pi}^2(y,k_\bot)-M_{0,K\pi}^2(x,q_\bot)
+ i\varepsilon} -\frac{1}{\mu^2-M_{ 0,K\pi}^2(x,q_\bot)}\right]
\label{xi0}
\end{eqnarray}
with the $K\pi$ system free squared-mass given by
$M_{ 0,K\pi}^2(x,q_\bot) = \frac{q_\bot^2 + m_\pi^2}{x} +
\frac{q_\bot^2 + m_K^2}{1-x}.$
After integration over $\theta$ and $q_\perp$, Eq.~(\ref{xi0}) is finally 
written as
\begin{eqnarray}
\xi_0(y,k_\bot)=\lambda(\mu^2)+ \frac{i}{4}\int_0^1 \frac{dx}{(2\pi)^2} \ln 
\frac{(1-x)(xM_{K\pi}^2(y,k_\bot)-m_\pi^2+ix\varepsilon)-xm_K^2}{
(1-x)(x\mu^2-m_\pi^2)-xm_K^2}.
\label{xi01}
\end{eqnarray}

\section{Application in the $B^+\to K^-\pi^+\pi^+$ decay}
\label{LFMDDecay}

The model for the $B^+\to K^-\pi^+\pi^+$ decay with FSI is based on an 
inhomogeneous integral equation for the spectator meson, with the meson-meson 
scattering amplitude as input. Isospin states of the $\pi\pi$ 
interaction are disregarded here, unlike the $I_{K\pi}=1/2$ and $I_{K\pi}=3/2$ states for 
the $K^\mp\pi^\pm$ channel, consider in our calculations. Our parametrization for the 
$K\pi$ amplitude follows the experimental results of \cite{LASS}, where the resonant 
$I_{K\pi}=1/2$ channel below $K^*_0(1430)$ dominates and the $I_{K\pi}=3/2$ amplitude is 
comparable. This model is the same used in Ref.~\cite{JHEP14} to study the $D^+\to 
K^-\pi^+\pi^+$ decay. A calculation up to two loops for this same decay was performed in 
Ref.~\cite{MagPRD11} bellow $K^*_0(1430)$. 

Here the LF model is applied to the $B$ decay and the calculations are 
performed up to three-loops in order to check the numerical convergence of 
the integrals. There are two possible total isospin states, namely, $I_T = 5/2$ 
and $3/2$. In our notation, the bachelor amplitude has the total isospin index and the 
one related with the interacting pair $\xi^{I_T^z}_{I_T,I_{K\pi}}(y,k_\bot)$, where we 
also consider the isospin projection index. The source amplitude written in terms of the 
$K\pi$ isospin state reads
\begin{align}
\left|B_0\right> = \sum_{I_T,I_{K\pi}}\alpha^{I_T^z}_{I_T,I_{K\pi}}
\left|I_T,I_{K\pi},I_T^z\right> + \sum_{I_T,I_{K\pi^\prime}}\alpha^{I_T^z}_{I_T,I_{
K\pi^\prime}}\left|I_T,I_{K\pi^\prime},I_T^z\right>,
\label{initial}
\end{align}
which has no dependence on the momentum variables and has an arbitrary 
normalization, since we are not considering explicitly short-distance processes in our 
calculations. For sake of simplicity we define the recoupling coefficients as 
$R^{I_T^z}_{I_T,I_{K\pi},I_{K\pi^\prime}}=\left<I_T,I_{K\pi},I_T^z|I_T,I_{
K\pi^\prime},I_T^z\right>$. This allows us to write the set of isospin coupled integral 
equations as
\begin{multline}
\xi^{I_T^z}_{I_T,I_{K\pi}}(y,k_\bot)=\left<I_T,I_{K\pi},I_T^z|B\right>
\xi_0(y,k_\bot)
+\frac{i}{2 }\sum_{I_{K\pi^\prime}}R^{I_T^z}_{I_T,I_{K\pi},I_{K\pi^\prime}}
\int_0^{1-y}\frac{dx}{x(1-y-x)}\int_0^\infty \frac{dq_\bot  }{(2\pi)^3} \\  \times
 K_{I_{K\pi^\prime}}(y,k_\bot;x,q_\bot)\,\xi^{I_T^z}_{I_T,I_{K\pi^\prime}}(x,q_\bot), \label{quilfisos}
\end{multline}
where the free squared mass of the $K\pi\pi$  system is
\begin{eqnarray}
M_{0,K\pi\pi}^2(x,q_\bot,y,k_\bot) = \frac{k_\bot^2 + m_\pi^2}{y}
+ \frac{q_\bot^2 + m_\pi^2}{x}
+ \frac{q_\bot^2 + k_\bot^2 + 2q_\bot k_\bot \cos\theta+ m_K^2}{1-x-y},
\end{eqnarray}
with the squared-mass of the virtual $K\pi$ system
$M_{K\pi}^2(z,p_\bot) =(1-z)\left(M_B^2 - \frac{p_\bot^2 +
m_\pi^2}{z}\right) - p_\bot^2. \,\,\,
$
The kernel carrying the $K\pi$ scattering amplitude is
\begin{eqnarray}
K_{I_{K\pi^\prime}}(y,k_\bot;x,q_\bot)=\int_0^{2\pi} d\theta \,\,
{q_\bot\,\tau_{I_{K\pi^\prime}}\left(M_{K\pi^\prime}^2(x,q_\bot)\right)\over M_B^2-M_{0,K\pi\pi}^2(x,q_\bot,y,k_\bot)+i\varepsilon}.
\label{kisos}
\end{eqnarray}
Isospin $2$ states of pion-pion interactions are not considered in the model, which 
will be explored as a single channel model, with the $K\pi$ s-wave interaction in the 
resonant $I=1/2$, and as a coupled channel model with both $I=1/2$ and $3/2$ $K\pi$ s-wave 
interactions.

The symmetrized decay amplitude with respect to the identical pions is written as
\begin{eqnarray}
{\cal A}_0 = A_0(M^2_{K\pi^\prime} ) +
A_0(M^2_{K\pi } ). \label{boseamplf}
\end{eqnarray}
The isospin projection on each term leads to
\begin{eqnarray}
A_0(M^2_{K\pi^\prime})
&=&\sum_{I_T,I_{K\pi^\prime},I_T^z}\left<K^-\pi^+\pi^+\right|
\left.I_T,I_{K\pi^\prime},I_T^z\right> \left[\frac{1}{2}\left<I_T,I_{K\pi^\prime},I_T^z\right|
\left.B_0\right>+\tau_{I_{K\pi }}(M^2_{K\pi^\prime})\xi^{I_T^z}_{I_T,I_{K\pi^\prime}}
(k_{\pi})\right] \nonumber \\ &=&a_0(M^2_{K\pi^\prime})e^{i \Phi_0(M^2_{K\pi^\prime})}.
\label{kpiamplf}
\end{eqnarray}

\section{Numerical perturbative solutions}
\label{pertsol}

The problem is solved by integrating the terms starting from the driving term and 
iterating as a perturbative series. The integrations are done up to three loops in order 
to check the convergence. In the coupled-channel calculations, the total isospin states 
$I=3/2$ are performed coupling $I_{K\pi}=1/2$ or $I_{K\pi}=3/2$ states. We also 
consider the $I_T=5/2$ with its single contribution in the $K\pi$ interaction for the 
isospin $3/2$ states. 

For the single channel case we consider only $K\pi$ interaction
in the resonant isospin 1/2 states and the perturbative solution of the equation
up to three-loops reduces to
\begin{multline}
\xi^{\mbox{\tiny{3/2}}}_{\mbox{\tiny{3/2,1/2}}}(y,k_\bot) =
\frac{1}{6}\sqrt{\frac{2}{3}}\xi_0(y,k_\bot) -
\frac{i}{3 }\left(\frac{1}{6}\sqrt{\frac{2}{3}}\right)
\int_0^\infty \frac{dq_\bot}{(2\pi)^3}\int_0^{1-y}dx\,K_{\mbox{\tiny 1/2}}(y,k_\bot;x,q_\bot)
\,\xi_0(x,q_\bot)+\\ -
\frac{1}{9}\left(\frac{1}{6}\sqrt{\frac{2}{3}}
\right)\int_0^\infty \frac{dq_\bot}{(2\pi)^3}\int_0^{1-y}dx\,
K_{\mbox{\tiny 1/2}}(y,k_\bot;x,q_\bot)
\int_0^\infty \frac{dq_\bot'}{(2\pi)^3}\int_0^{1-x}dx'\,K_{\mbox{\tiny
1/2}}(x,q_\bot;x',q_\bot')\,\xi_0(x',
q_\bot')+\cdots
\label{itersol-single}
\end{multline}
where we compute driving term considering $\alpha^{3/2}_{3/2,1/2}=1$ and the kernel 
$K_{\mbox{\tiny 1/2}}$ is defined by Eq.~(\ref{kisos}). The factor $\mathcal{N}$ is a 
normalization representing the source partonic amplitudes and is assumed constant.

The numerical integration over the radial variable is computed introducing a momentum 
cut-off $\Lambda=0.8$~GeV. This is smaller than in the $D$ decay case, but in that case 
the change of the cutt-off parameter from $2.0$~GeV to $0.8$~GeV practically 
does not alters the results. In the $B$ case, the use of 
$\Lambda=2.0$~GeV is very expensive numerically, and probably this is related to the 
large non-physical region accessed. Since the $B$ meson is much heavier than the $D$ 
one, its wave function is concentrated in the low-momentum region. The partonic decay 
amplitudes should also carry this behavior and, since we are not computing the 
amplitudes at this level, we simulate this effect through the momentum 
cut-off parameter. We address a detailed analysis on this issue to future 
works.

For the numerical calculations, a finite value for $\varepsilon$ was used in the meson 
propagators, in order to take into account the absorption related to different possible 
decay channels. The value used here was $\varepsilon=0.5$~GeV$^2$, which is larger than 
the one used in the $D$ decay case. In fact, since the $B$ phase space is very large, we 
know that the absorption is higher comparing with the $D$ decay. Here we mimic this  
effect by using a larger value for the $\varepsilon$ parameter. We have tested different 
values of $\varepsilon$ (close to $\varepsilon=0.5$~GeV$^2$), obtaining a small 
difference in the results. The subtraction constant in the driving term is chosen to be 
zero.

\begin{figure}[!htb]
\centering
\includegraphics[scale=0.37]{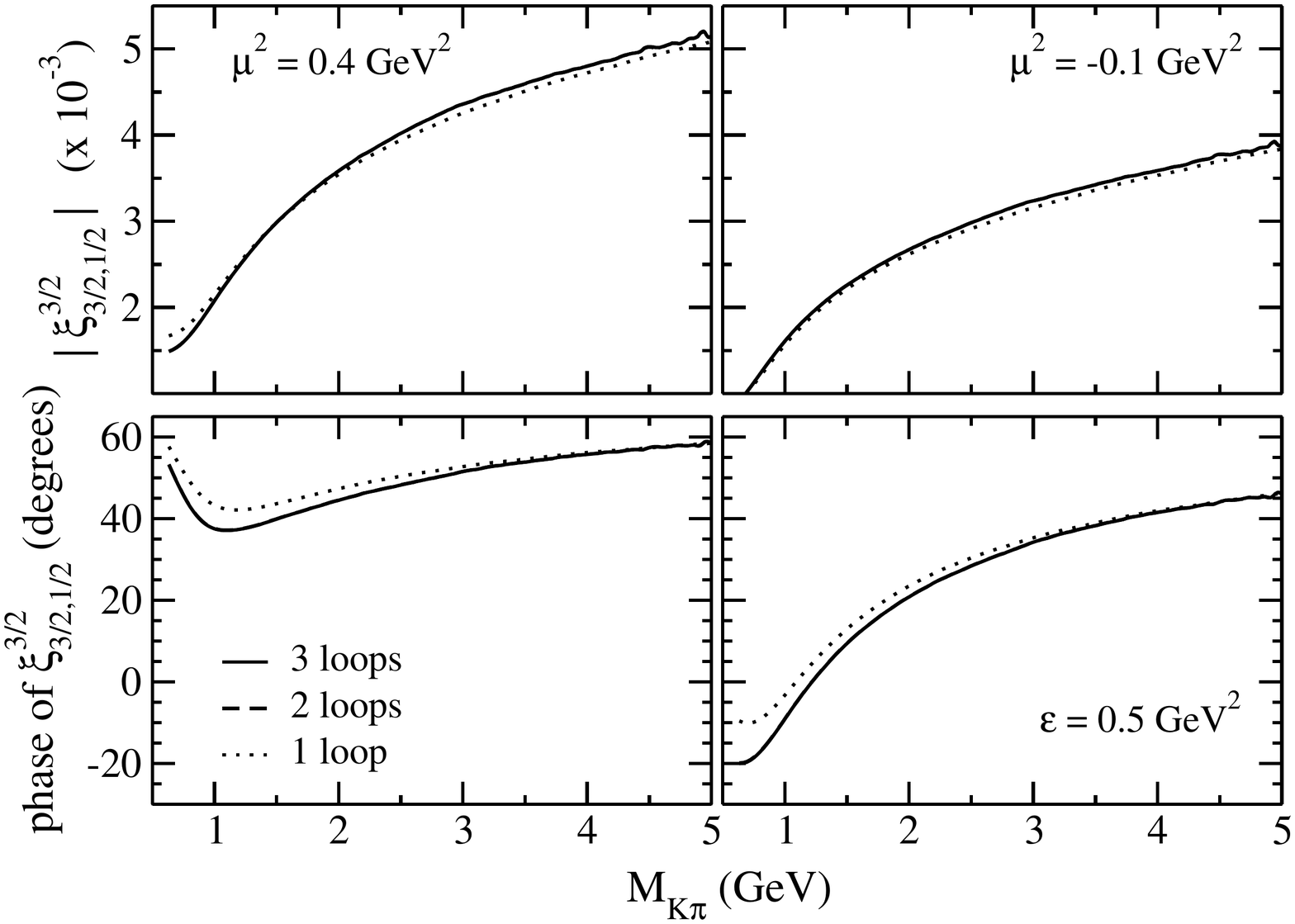}
\caption{Modulus and phase of $\xi^{\mbox{\tiny{3/2}}}_{\mbox{\tiny{3/2,1/2}}}$
for $\varepsilon=0.5$ GeV$^2$, $\mu^2=0.4$~GeV$^2$ (left) and $\mu^2=-0.1$~GeV$^2$ 
(right).}
\label{qsisingle}
\end{figure}

Regarding the convergence of the loop expansion, we have studied it up 
to three-loops. The results concerning phase and modulus of the bachelor function is 
depicted in Fig.~\ref{qsisingle}. 

It was used $\mu^2=(0.4,-0.1)$~GeV$^2$, in order to verify the effect of 
the subtraction point in the calculations. We have also used 
$\varepsilon=0.5$ GeV$^2$. For a fixed value of $\mu^2$ it is clear that the 
two-loop solution already present convergence and is enough for practical 
applications. This finding is similar to that observed in the $D$ 
decay case, but now the results are even better concerning the convergence. 
The phase is always positive for $\mu^2=0.4$~GeV$^2$, but can be either positive or 
negative for $\mu^2=-0.1$~GeV$^2$. The phase variation is large for $\mu^2=-0.1$~GeV$^2$ 
and presents a minimum increasing again for $\mu^2=0.4$~GeV$^2$. In both cases the modulus 
increases for larger two-body invariant masses.

\subsection{Interaction for coupled channels in $I_{K\pi}=1/2$ and 3/2 states}

In the coupled channels case the set of integral equations obtained from Eq.~(\ref{quilfisos}) reads
\begin{multline}
\xi^{\mbox{\tiny{3/2}}}_{\mbox{\tiny{3/2,1/2}}}(y,k_\bot)=A_w\,\xi_0(y,k_\bot)
+\frac{iR^{\mbox{\tiny{3/2}}}_{\mbox{\tiny{3/2,1/2,1/2}}}}{2}\int_0^{1-y
}\frac{dx}{x(1-y-x)}\int_0^\infty \frac{dq_\bot}{(2\pi)^3} K_{\mbox{\tiny 1/2}}(y,k_\bot;x,q_\bot)\,\xi^{\mbox{\tiny{3/2}}
}_{\mbox{\tiny{3/2,1/2}}}(x,q_\bot) + \\
+\frac{iR^{\mbox{\tiny{3/2}}}_{\mbox{\tiny{3/2,1/2,3/2}}}}{2(2\pi)^3}
\int_0^{1-y}\frac{dx}{x(1-y-x)}\int_0^\infty \frac{dq_\bot}{(2\pi)^3}\,
K_{\mbox{\tiny 3/2}}(y,k_\bot;x,q_\bot)\,\xi^{\mbox{\tiny{3/2}}
}_{\mbox{\tiny{3/2,3/2}}}(x,q_\bot),
\label{eqint-coupled1}
\end{multline}
\begin{multline}
\xi^{\mbox{\tiny{3/2}}}_{\mbox{\tiny{3/2,3/2}}}(y,k_\bot)=B_w\, \xi_0(y,k_\bot)
+\frac{iR^{\mbox{\tiny{3/2}}}_{\mbox{\tiny{3/2,3/2,1/2}}}}{2}\int_0^{1-y
}\frac{dx}{x(1-y-x)}\int_0^\infty \frac{dq_\bot}{(2\pi)^3}\,
K_{\mbox{\tiny 1/2}}(y,k_\bot;x,q_\bot)\,\xi^{\mbox{\tiny{3/2}}
}_{\mbox{\tiny{3/2,1/2}}}(x,q_\bot)   + \\
+\frac{iR^{\mbox{\tiny{3/2}}}_{\mbox{\tiny{3/2,3/2,3/2}}}}{2}
\int_0^{1-y}\frac{dx}{x(1-y-x)}\int_0^\infty \frac{dq_\bot}{(2\pi)^3}\,K_{\mbox{\tiny 3/2}}(y,k_\bot;x,q_\bot)\,\xi^{\mbox{\tiny{3/2}}
}_{\mbox{\tiny{3/2,3/2}}}(x,q_\bot) .
\label{eqint-coupled2}
\end{multline}
and for $I_T=5/2$
\begin{multline}
\xi^{\mbox{\tiny{3/2}}}_{\mbox{\tiny{5/2,3/2}}}(y,k_\bot)=C_w\,\xi_0(y,k_\bot)
+\frac{iR^{\mbox{\tiny{3/2}}}_{\mbox{\tiny{5/2,3/2,3/2}}}}{2 }\int_0^{1-y
}\frac{dx}{x(1-y-x)}\int_0^\infty \frac{dq_\bot}{(2\pi)^3}\,K_{\mbox{\tiny 3/2}}(y,k_\bot;x,q_\bot)
\, \xi^{\mbox{\tiny{3/2}}}_{\mbox{\tiny{5/2,3/2}}}(x,q_\bot),
\label{eqint-coupled3}
\end{multline}
where the isospin states related to the projection of the partonic amplitude 
(\ref{initial}) brings the weights $A_w$, $B_w$ and $C_w$, given by
$A_w=\left<I_T=3/2,I_{K\pi}=1/2,I_T^z=3/2\right|\left.B_0\right>/2$, 
$B_w=\left<I_T=3/2,I_{K\pi}=3/2,I_T^z=3/2\right|\left.B_0\right>2/$ and
$C_w=\left<I_T=5/2,I_{K\pi}=3/2,I_T^z=3/2\right|\left.B_0\right>/2$
where the isospin coefficients are
$A_w =
\alpha^{\mbox{\tiny{3/2}}}_{\mbox{\tiny{3/2,1/2}}}
(1+R^{\mbox{\tiny{3/2}}}_{\mbox{\tiny{3/2,1/2,1/2}}})
+ \alpha^{\mbox{\tiny{3/2}}}_{\mbox{\tiny{3/2,3/2}}}
R^{\mbox{\tiny{3/2}}}_{\mbox{\tiny{3/2,1/2,3/2}}}$,
$B_w = \alpha^{\mbox{\tiny{3/2}}}_{\mbox{\tiny{3/2,3/2}}}
(1+R^{\mbox{\tiny{3/2}}}_{\mbox{\tiny{3/2,3/2,3/2}}})
+ \alpha^{\mbox{\tiny{3/2}}}_{\mbox{\tiny{3/2,1/2}}}
R^{\mbox{\tiny{3/2}}}_{\mbox{\tiny{3/2,3/2,1/2}}}$ and
$C_w = \alpha^{\mbox{\tiny{3/2}}}_{\mbox{\tiny{5/2,3/2}}}
(1+R^{\mbox{\tiny{3/2}}}_{\mbox{\tiny{5/2,3/2,3/2}}}).$
The coefficients $\alpha$ come from the partonic decay amplitude (\ref{initial}) 
projected onto the isospin space and are defined as
$\alpha^{\mbox{\tiny{3/2}}}_{\mbox{\tiny{3/2,1/2}}}= W_1\,
C^{\mbox{\tiny{1/2  1   3/2}}}_{\mbox{\tiny{1/2  1   3/2}}} \,
C^{\mbox{\tiny{1   1/2  1/2}}}_{\mbox{\tiny{1  -1/2  1/2}}}/2$,
$\alpha^{\mbox{\tiny{3/2}}}_{\mbox{\tiny{3/2,3/2}}}=W_2\,
C^{\mbox{\tiny{3/2  1   3/2}}}_{\mbox{\tiny{1/2   1   3/2}}}\,
C^{\mbox{\tiny{1   1/2  3/2}}}_{\mbox{\tiny{1   -1/2  1/2}}}/2$ and 
$\alpha^{\mbox{\tiny{3/2}}}_{\mbox{\tiny{5/2,3/2}}}=W_3\,
C^{\mbox{\tiny{3/2  1  5/2}}}_{\mbox{\tiny{1/2 1  3/2}}}\,
C^{\mbox{\tiny{1  1/2  3/2}}}_{\mbox{\tiny{1 -1/2  1/2}}}/2$
with the parameters $W_1=W_2=W_3=1$ for the case 
$\left|B_0\right>=\left|K^-\pi^+\pi^+\right>$  and
the Clebsch-Gordan and recoupling
coefficients 
$\,C^{\mbox{\tiny{1/2  1   3/2}}}_{\mbox{\tiny{1/2  1   3/2}}}=1\,$,
$\,C^{\mbox{\tiny{1   1/2  1/2}}}_{\mbox{\tiny{1  -1/2  1/2}}}=\sqrt{2/3}\,$,
$\,C^{\mbox{\tiny{3/2  1   3/2}}}_{\mbox{\tiny{1/2   1   3/2}}}=-\sqrt{2/5}\,$,
$\,C^{\mbox{\tiny{1   1/2  3/2}}}_{\mbox{\tiny{1   -1/2  1/2}}}=1/\sqrt{3}\,$,
$\,C^{\mbox{\tiny{3/2  1  5/2}}}_{\mbox{\tiny{1/2 1  3/2}}}=\sqrt{3/5}\,$,
$\,R^{\mbox{\tiny{3/2}}}_{\mbox{\tiny{3/2,1/2,1/2}}}=-2/3\,$,
$\,R^{\mbox{\tiny{3/2}}}_{\mbox{\tiny{3/2,1/2,3/2}}}=\sqrt{5}/3\,$,
$\,R^{\mbox{\tiny{3/2}}}_{\mbox{\tiny{3/2,3/2,3/2}}}=2/3\,$,
$\,R^{\mbox{\tiny{3/2}}}_{\mbox{\tiny{3/2,3/2,1/2}}}=\sqrt{5}/3\,$, and
$\,R^{\mbox{\tiny{3/2}}}_{\mbox{\tiny{5/2,3/2,3/2}}}=1\,$.
With all these manipulations the weights $A_w$, $B_w$,
and $C_w$ reads
$A_w = \sqrt{\frac{1}{54}}(W_1-W_2)$,
$B_w = \sqrt{\frac{5}{54}}(W_1-W_2)$ and
$C_w = \frac{W_3}{\sqrt{5}}$.

\begin{figure}[!htb]
\centering
\includegraphics[scale=0.42]{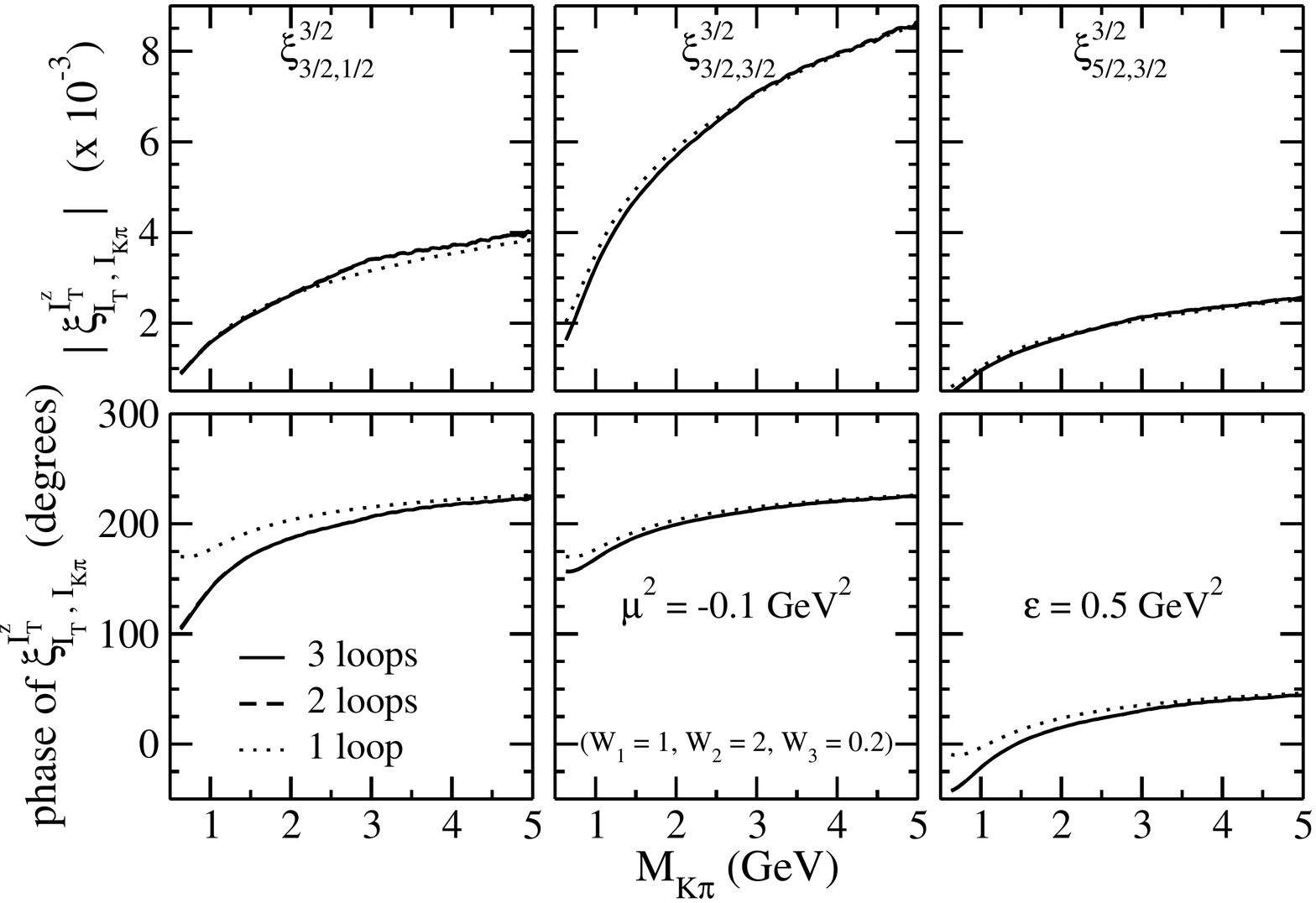}
\caption{Modulus and phase of $\xi^{I_T^z}_{I_T,I_{K\pi}}$
for $\varepsilon=0.5$ GeV$^2$ and $\mu^2=-0.1$ GeV$^2$ and the parameters $W_1=1$, $W_2=2$ 
and $W_3=0.2$.}
\label{qsicoupled}
\end{figure}

Also in this coupled channels case, the bachelor amplitude is computed to check the 
convergence. The coupled equations of Eq.~(\ref{eqint-coupled1}) appear in the case 
$I_T=3/2$. For $I_T=5/2$, it is a single channel equation Eq.~(\ref{eqint-coupled3}). The 
results are shown in Fig.~\ref{qsicoupled}, using $\varepsilon=0.5$ GeV$^2$ and 
$\mu^2=-0.1$~GeV$^2$, with the parameters from the expansion of the source term  
given by $W_1=1$, $W_2=2$ and $W_3=0.2$. 

Again, the convergence is clear and the two-loop result is already enough for practical 
applications. Both, phase and modulus of the bachelor amplitudes increases with 
$M_{K\pi}$ and changes considerably along the large phase space available. In the channel 
$I_T=3/2$, both components have similar magnitudes for the phase and are larger than  
that from the $I_T=5/2$ case, same pattern observed in the $D$ decay.

\section{Results for the Phase and Amplitude in the $B^+\to K^-\pi^+\pi^+$ decay}

Since the two-loop result presents already a good convergence for the bachelor 
amplitudes, we restrict our calculations hereafter to decay amplitude up to 
two-loops in Eq.~(\ref{itersol-single}). For the moment, there is no experimental 
data available to perform a comparative analysis as done for the $D$ meson decay 
in~\cite{JHEP14}. For the single channel calculations we consider only the S-wave $K\pi$ 
scattering amplitude in the isospin $1/2$ state, which is fitted to the LASS 
data~\cite{LASS}. The reduced form of the decay amplitude, that will give us both phase 
and modulus by means of Eq.~(\ref{kpiamplf}), reads
\begin{align}
A_0(M^2_{K\pi})
=\sqrt{\frac{2}{3}}\left[\frac{1}{12}\sqrt{\frac{2}{3}} +
\tau_{\mbox{\tiny{1/2}}}(M^2_{K\pi})\xi^{\mbox{\tiny{3/2}}}_
{\mbox{\tiny{3/2,1/2}}}(k_{\pi^\prime})\right].
\label{amplitude}
\end{align}

The iteration of the coupled equations (\ref{eqint-coupled1})-(\ref{eqint-coupled2}) 
gives the results for the channel $I_T=3/2$. For the the $I_T=5/2$ state, 
the amplitude is given by the single expression in Eq.~(\ref{eqint-coupled3}). We 
also consider for these calculations the results up to two loops, since the convergence 
is verified. The S-wave decay amplitude is
\begin{eqnarray}
A_0(M^2_{K\pi})&=&C_1\left[\frac{A_w}{2}
+\tau_{\mbox{\tiny{1/2}}}(M^2_{K\pi})\xi^{\mbox{\tiny{3/2}}}
_{\mbox{\tiny{3/2,1/2}}}(k_{\pi^\prime})\right]
+C_2\left[\frac{B_w}{2}
+\tau_{\mbox{\tiny{3/2}}}(M^2_{K\pi})\xi^{\mbox{\tiny{3/2}}}
_{\mbox{\tiny{3/2,3/2}}}(k_{\pi^\prime})\right] + \nonumber \\
&+&C_3\left[\frac{C_w}{2}
+\tau_{\mbox{\tiny{3/2}}}(M^2_{K\pi})\xi^{\mbox{\tiny{3/2}}}
_{\mbox{\tiny{5/2,3/2}}}(k_{\pi^\prime})\right]
\label{amplitude-coupled}
\end{eqnarray}
where the constants $C_i$ come from the isospin projection onto the state 
$K\pi\pi$, Eq.~(\ref{kpiamplf}). There are two free parameters related with the projected 
partonic amplitude, namely, $W_1-W_2$ and $W_3$. If the first is zero and the second 
nonzero, only total isospin $5/2$ appears and there is no structure in the decay 
amplitude, as shown in Ref.~\cite{JHEP14}. This shows that it is not a good physical 
solution, since the isospin state contributions are not being taken into account in 
a reasonable way. A more detailed study of the correct weights using the LF model would 
be guided by experimental data, as done for the $D^+\to K^-\pi^+\pi^+$ decay 
in~\cite{JHEP14}. Here we just follow that study, where the authors found a small 
mixture of the total isospin $5/2$ state.

\begin{figure}[!htb]
\centering
\includegraphics[scale=0.42]{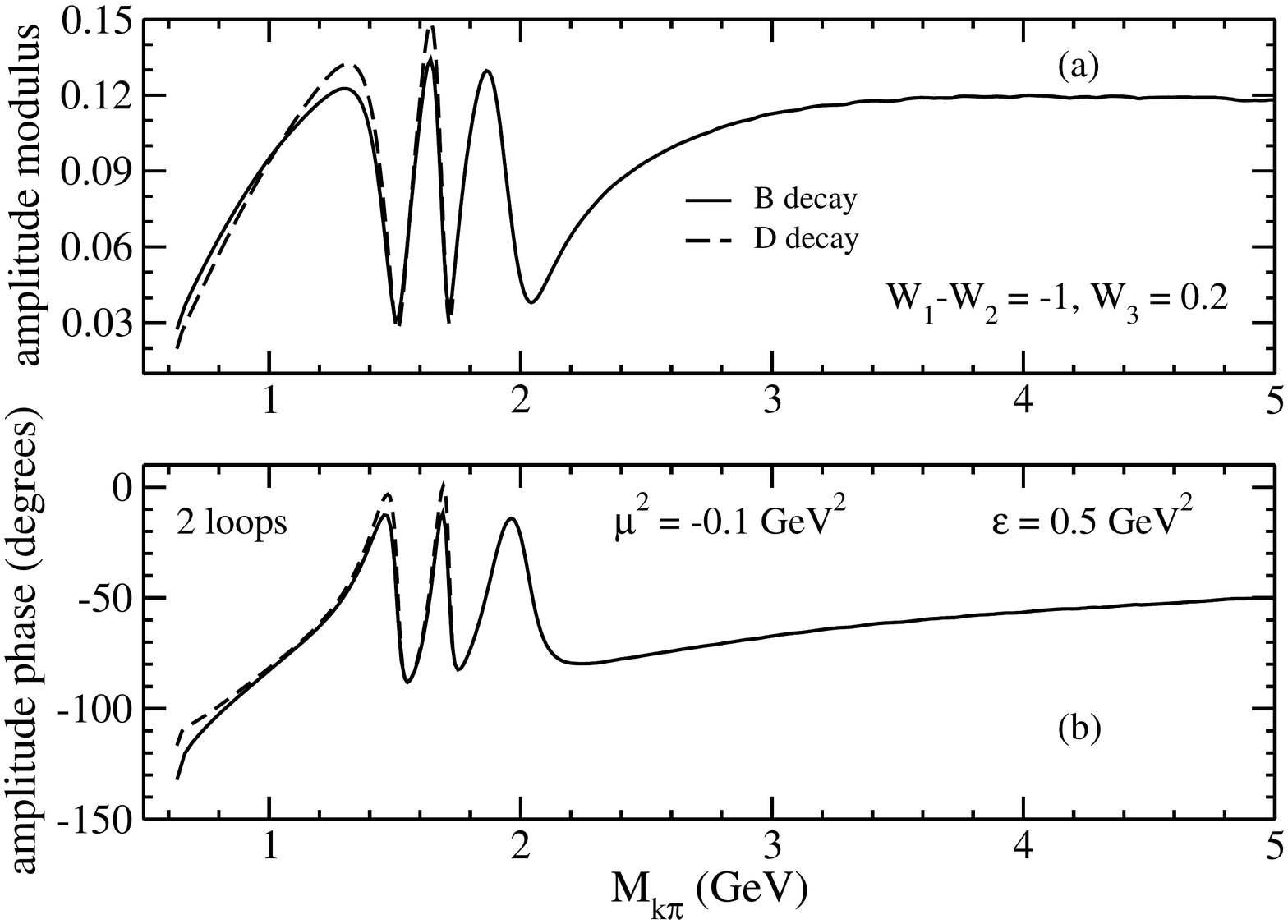}
\caption{Comparison of (a) modulus and (b) phase between $D^+\to K^-\pi^+\pi^+$ and 
$B^+\to K^-\pi^+\pi^+$ amplitudes for a initial state in which $W_1-W_2=-1$ and 
$W_3=0.2$.}
\label{DBcomp}
\end{figure}

In Fig.~\ref{DBcomp} we show a comparison between modulus and phase of decay 
amplitudes for the $B^+$ and $D^+$ mesons, both decaying to the same final state 
$K^-\pi^+\pi^+$. 
The subtraction scale is fixed in $\mu^2=-0.1$~GeV$^2$, the  
$\varepsilon$ parameter was chosen to be $\varepsilon=0.5$~GeV$^2$, and $W_1-W_2=-1$ and 
$W_3=0.2$ were used. All these parameters are kept the same for both 
cases. In order to test the effect of the constants $W_1-W_2$ and $W_3$, we 
have tried a second set of parameters, namely, $W_1-W_2=1$ and $W_3=0.3$, which was 
used in Ref.~\cite{JHEP14} to study the experimental data for the $D^+ \to K^- \pi^+ 
\pi^+$ decay amplitude, but the results are very similar and with only a change of 
sign in the phase.

\begin{figure}[!htb]
\centering
\includegraphics[scale=0.42]{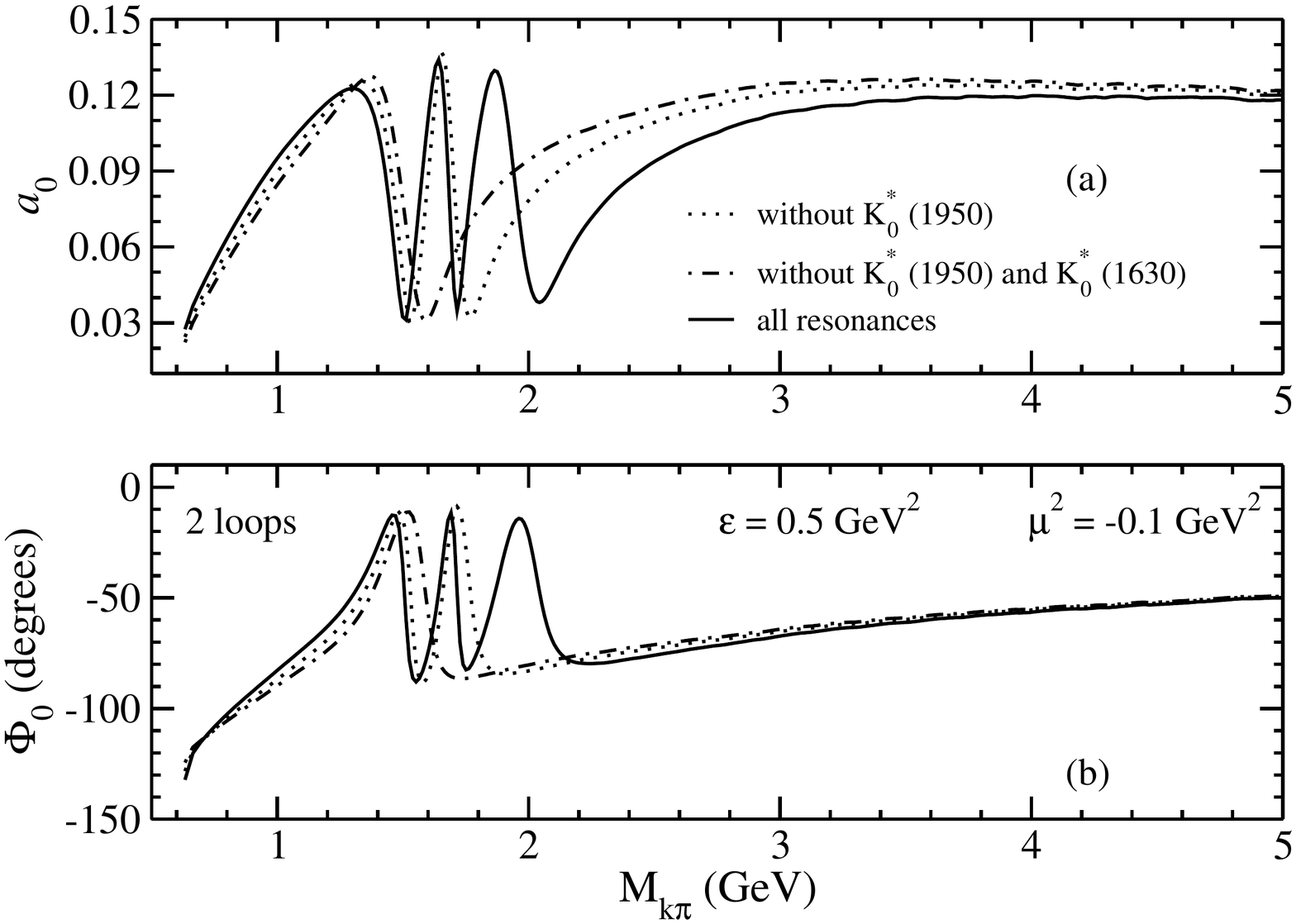}
\caption{ Modulus ($a_0$) and phase ($\Phi_0$) of the $B^+\to K^-\pi^+\pi^+$ 
amplitude, in the notation of Eq.~(\ref{kpiamplf}), comparing cases with all resonances, 
without $K^*_0(1950)$ and without both $K^*_0(1950)$ and $K^*_0(1630)$.}
\label{comp-res-nores}
\end{figure}

In Fig.~\ref{comp-res-nores} we compare both modulus and phase of the $B^+\to 
K^-\pi^+\pi^+$ decay amplitude with and without the resonant structure, which incorporates 
$K^*_0(1630)$ and $K^*_0(1950)$. 
For this study, we also fix the subtraction point at $\mu^2=-0.1$ GeV$^2$ and the other 
parameters with the same values as before. The figure shows that the inclusion of 
the resonances produces more bands in both modulus and phase. This is clearly related with 
the resonances, since the peaks are around its masses and bellow $K^*_0(1430)$ the effect 
is small. All the cases have the same tail when the two-body invariant mass 
increases. In amplitude analysis of the three-body $B$ decay to the $K\pi\pi$ channel the 
$K_0^*(1630)$ resonance is usually included explicitly in the fit, insofar the 
$K_0^*(1950)$ is more complicated to claim that exists in the channel, therefore it 
appears indirectly in experimental analysis.

\section{Summary and Conclusions}
\label{conclusion}

In this work we have used a Light-front framework to compute off-shell decay amplitudes 
starting from the four-dimension Bethe-Salpeter equation decomposed in the Faddeev form. 
The contribution of final state interactions to the $B^+\to K^-\pi^+\pi^+$ decay is 
obtained. This approach can be applied for charged three-body heavy meson decays, and was 
used before for the $D$ meson decay, and the calculations were compared to the 
experimental data expressed in terms of the modulus and phase-shift~\cite{JHEP14}. 
Here, we have used the same three-body rescattering model in the final state for the $B 
\to K \pi \pi$ decay, considering the S-wave $K\pi$ interactions in the resonant $1/2$ 
state, the $K^*_0(1430)$, $K^*_0(1630)$ and $K^*_0(1950)$ resonances and the non-resonant 
$3/2$ isospin states. The scattering matrix was parametrized and fixed with the 
requirement of fitting the LASS data~\cite{LASS}, as done in the $D$ decay 
case~\cite{JHEP14}.

In the Light-front, the inhomogeneous integral equations reduce to three-dimensional ones, 
solved here with a perturbative series up to three-loops and with the accuracy of the 
solution checked. The convergence of the series is clear and the two-loop results shows up 
enough for practical applications, as happened in the $D$ meson decay case of 
Ref.~\cite{JHEP14}. In comparison with the decay of the lighter $D$ meson, we needed to 
use a larger imaginary part for the propagators of the mesons by increasing the 
$\varepsilon$ parameter. Since this parameter mimics the absorption to other decay 
channels, it is expected that in the $B$ decay, $\varepsilon$ increases due to the much 
larger phase space available. The momentum cut-off was chosen smaller in the $B$ case than 
in the $D$ decay, in order to have a good convergence. Such a decrease seems reasonable as 
$B$ is much more massive than $D$. The heavier particle should have a larger number of 
decay channels, meaning larger absorption, and therefore the wave function of the 
particular decay channel at short-distances, where the absorption takes place, is 
suppressed. The result is that the outgoing state is more concentrated at large distances, 
which corresponds to the low-momentum region. The smaller cut-off in the $B$ decay with 
respect to the $D$ one, can be understood as an effective way to parametrize the physics 
of the larger number of open channels.

The resonant structure above the $K_0^*(1430)$ resonance is also a question that 
deserves a detailed analysis in face of future experimental data. While the 
presence of the $K_0^*(1630)$ resonance is expected, and this is in fact used in 
our amplitude analysis, the $K_0^*(1950)$ influence must be better understood. 
Other aspect that requires more study are the real weights of the three isospin 
components of the source amplitudes at the quark level. Three-body rescattering effects 
are also important because they distribute $CP$ violation to different decay channels, 
since it is one of the mechanisms allowed by the $CPT$ constraint~\cite{BnoCPV}. In the 
near future, this light-front approach will  be generalized in order to study $CP$ 
violation in three-body charmless $B$ decays, taking into account the unitarity of the 
S-matrix, and the $CPT$ constraint, exactly as done in Refs.~\cite{CPT,Nogueira15}.

\acknowledgments

We thank the Brazilian funding agencies Funda\c{c}\~{a}o de Amparo \`a Pesquisa do Estado 
de S\~{a}o Paulo (FAPESP) and Conselho Nacional de Desenvolvimento Cient\'ifico e 
Tecnol\'ogico (CNPq). J.H.A.N. also acknowledges the support of Grant No. 2014/19094-8 
from FAPESP.

\section*{Competing Interests}
The authors declare that there is no conflict of interest regarding the publication of 
this paper.


\end{document}